\documentclass{PoS}

\title{First Results from HAWC on GRBs}

\ShortTitle{GRBs with HAWC}

\author{\speaker{Dirk Lennarz}\thanks{now at: Department of Physics \& Astronomy, Michigan State University, East Lansing, MI, USA.} $^{a}$, Ignacio Taboada$^{a}$ for the HAWC collaboration$^{b}$\\
        \llap{$^a$}School of Physics and Center for Relativistic Astrophysics, Georgia Institute of Technology\\
                   Atlanta, Georgia, USA\\
        \llap{$^b$}For a complete author list, see \href{http://www.hawc-observatory.org/collaboration/icrc2015.php}{www.hawc-observatory.org/collaboration/icrc2015.php}.\\
        Email: \email{dirk.lennarz@gatech.edu}}

\abstract{In this contribution, the first results of HAWC, searching for VHE gamma-ray emission from gamma-ray bursts (GRBs) reported by \emph{Swift}, are presented. The HAWC gamma-ray observatory is operating in central Mexico at an altitude of 4,100~m above sea level. With an instantaneous field of view of approximately $2~\rm{sr}$ and over 95\% duty cycle (up time fraction), HAWC is an ideal detector to perform ground-based gamma-ray observations of GRBs. Though optimised for TeV observations, HAWC has significant sensitivity to short transients of energies as small as 50~GeV. The analysis method used for fast online and offline HAWC follow up of GRBs reported by satellites is described.}


\FullConference{The 34th International Cosmic Ray Conference,\\
		30 July- 6 August, 2015\\
		The Hague, The Netherlands}

\newcommand{\linebreakcell}[2][c]{%
  \begin{tabular}[#1]{@{}c@{}}#2\end{tabular}}

\newcommand\arcdeg{\mbox{$^\circ$}}

\begin{document}

\section{Introduction}
Gamma-ray bursts (GRBs) were discovered almost fifty years ago \cite{bib:GRB_review}, but despite thorough study, details of their particle acceleration mechanisms still remain in the dark. A recent observation of the Large Area Telescope (LAT) on board the \emph{Fermi Gamma-Ray Space Telescope} (\emph{Fermi}-LAT) has established that GRBs produce photons in the very-high-energy (VHE, $>100$~GeV) regime, when a $95.3~\rm{GeV}$ photon (or $128~\rm{GeV}$ when corrected for redshift) was detected in GRB~130427A \cite{bib:GRB130427A_Fermi_Science}. The effective area of satellite experiments is very limited at these energies and since the photon flux decreases steeply with energy, there is little hope of detecting more than just single photons. The highest energies and temporal evolution of GRB spectra carry important information about GRB physics, e.g. would the observation of VHE photons at late or early times challenge scenarios, in which those photons are produced by synchrotron emission. Furthermore, VHE observations can also provide insights into the extragalactic background light (EBL) and Lorentz invariance violation.

At VHEs, ground-based experiments might be able to supply those important measurements. Imaging Atmospheric Cherenkov Telescopes (IACTs) are pointed instruments that need to slew to the GRB position and will therefore in general miss the prompt and early afterglow phase. In contrast, the High Altitude Water Cherenkov (HAWC) observatory has a large instantaneous field of view ($\sim2~\rm{sr}$ or 16\% of the sky), $>95$\% duty cycle and the lack of observational delays will allow observing the GRB prompt phase. This makes HAWC an ideal detector for studying transient sources like GRBs.

\section{High Altitude Water Cherenkov Observatory}
HAWC is a new, ground-based VHE gamma-ray air-shower detector that recently finished construction at Sierra Negra, Mexico, at an altitude of 4,100~m above sea level \cite{bib:HAWC_highlights}. It applies the water Cherenkov technique, where VHE photons are detected by measuring Cherenkov light from secondary particles in an extensive air shower. The full detector consist of 300 steel tanks, containing light-tight bladders of 7.3~m diameter and 4.5~m depth, holding $\sim200,000$~litres of filtered water. The Cherenkov light is measured by three $8^{\prime\prime}$ photomultiplier tubes (PMTs) and one $10^{\prime\prime}$ PMT at the bottom of each tank. The PMT pulses are shaped and discriminated at two thresholds and the threshold crossing times are digitised. All signals are recorded to memory and triggering is entirely done in software by applying a threshold on the number of coincident PMT signals in a 150~ns time window (trigger threshold).

HAWC features two data acquisition (DAQ) systems. One is designed to read out full air-shower events (main DAQ), recording the time and charge of individual PMT pulses and the signal arrival time in different tanks. This makes it possible to reconstruct the incident direction of the shower. The dominant background when observing VHE gamma-rays comes from the abundant population of hadronic cosmic rays. A set of cuts can be used to remove hadronic events (gamma-hadron separation). The other DAQ (scaler DAQ) counts the signals in each PMT, allowing to detect GRBs by a statistical excess over the noise rate (``single particle technique'' \cite{bib:scaler_method}). Both DAQs have sensitivity to GRBs at different energies and therefore complement each other \cite{bib:HAWC_GRBs}. In this contribution, the analysis of the main DAQ data is described.

Science operations of the HAWC array started with a partially built array in August 2013 \cite{bib:Andy-ICRC-2015}. The stage of the detector between August 2, 2013 and July 8, 2014 is called HAWC-111. During this time, the uptime fraction was 83\%, which includes shutdowns to accommodate ongoing construction. The detector operated with a trigger threshold between 15 and 25 PMTs. The data presented in this contribution are from the HAWC-111 period.

Data was collected before the HAWC-111 period with a smaller detector. Into that period falls GRB~130427A, the most powerful burst ever detected with a redshift $z\lesssim0.5$. HAWC provided the first limits on the prompt emission of GRB~130427A in the VHE range, but due to the high zenith angle of the GRB and the incomplete HAWC detector, the limits are about two orders of magnitude higher than a simple extrapolation of the \emph{Fermi} data \cite{bib:HAWC_GRB130427A}.

\section{GRB Selection}
GRBs in the field of view during HAWC-111 operations were found using the LAT\footnote{http://swift.gsfc.nasa.gov/archive/grb\_table/}, \emph{Fermi} Gamma-ray Burst Monitor (\emph{Fermi}-GBM)\footnote{https://heasarc.gsfc.nasa.gov/W3Browse/all/fermigbrst.html} and \emph{Swift}\footnote{http://swift.gsfc.nasa.gov/archive/grb\_table/} online tables. The GRBOX: Gamma-Ray Burst Online Index\footnote{http://www.astro.caltech.edu/grbox/grbox.php} was searched without a match for additional GRBs from other experiments.

Considering GRBs down to a zenith angle of 51 degrees, there was only one LAT detected GRB in the HAWC field of view (GRB~130907A) at a zenith angle of $27\arcdeg$. Unfortunately, this burst occurred during one of the detector downtimes and no data is available. About 40 GBM bursts happened during HAWC-111 and for six of them there is no data available. Since October 2013, only one GBM burst has no data available, which demonstrates the excellent uptime fraction of HAWC. GBM-detected bursts are often poorly localised and require a different analysis technique than what is presented here, so this contribution focuses on the well localised \emph{Swift} bursts.

During HAWC-111, 22 \emph{Swift}-detected GRBs occurred in the HAWC field of view. Three of them (GRB~130907A, GRB~130912A and GRB~131018A) were missed due to detector downtime. GRB~140226A, which is the first burst discovered optically and independent of a high-energy trigger by the Intermediate Palomar Transient Factory \cite{bib:iPTF_GRB140226A}, peaked in Konus-\emph{Wind} at 10:02:57~UTC. Unfortunately, no data is available for the prompt phases of this GRB due to a run transition. Table~\ref{tab1} lists the 18 bursts analysed in this contribution.

\begin{table}
\footnotesize
\begin{tabular}{l|ccccccc}
\hline
GRB & \linebreakcell[t]{Trigger \\ Number} & \linebreakcell[t]{Time \\ UTC} & RA J2000 & DEC J2000 & \linebreakcell[t]{Zenith Angle \\ deg} & \linebreakcell[t]{BAT T90 \\ s} & \linebreakcell[t]{Significance \\ $\sigma$} \\
\hline
140628A & 602803 & 13:35:37 & 02h42m39.88s & -0d23m05.7s & 26.0 & 10.5 & -0.74 \\
140622A & 602278 & 09:36:04 & 21h08m41.56s & -14d25m09.5s & 33.4 & 0.13 & -0.93 \\
140607A & 601051 & 17:13:31 & 05h45m29.52s & 18d54m14.4s & 27.9 & 109.9 & 3.42 \\
140518A & 599287 & 09:17:46 & 15h09m00.60s & 42d25m05.6s & 48.6 & 60.5 & -0.61 \\
140430A & 597722 & 20:33:36 & 06h51m44.61s & 23d01m25.2s & 31.3 & 173.6 & -1.75 \\
140423A & 596901 & 08:31:53 & 13h09m08.54s & 49d50m29.4s & 46.9 & 134 & 0.21 \\
140419A & 596426 & 04:06:51 & 08h27m57.56s & 46d14m25.3s & 45.3 & 94.7 & 1.35 \\
140414A & GA     & 06:06:29 & 13h01m14.40s & 56d54m07.2s & 37.8 & 0.7 & -0.18 \\
140408A & 595141 & 13:15:54 & 19h22m51.83s & -12d35m42.5s & 32.4 & 4.00 & -0.02 \\
140331A & 594081 & 05:49:48 & 08h59m27.46s & 02d43m02.3s & 45.7 & 209 & -2.18 \\
140215A & 586680 & 04:07:10 & 06h56m35.81s & 41d47m11.7s & 23.2 & 84.2 & 0.30 \\
140206A & 585834 & 07:17:20 & 09h41m20.26s & 66d45m38.6s & 47.7 & 93.6 & -1.86 \\
140129A & 585128 & 03:23:59 & 02h31m33.78s & -01d35m43.4s & 47.8 & 2.99 & 1.65 \\
140114A & 583861 & 11:57:40 & 12h34m05.16s & 27d57m02.6s & 11.1 & 139.7 & 0.29 \\
131229A & 582374 & 06:39:24 & 05h40m55.61s & -04d23m46.7s & 27.7 & 13.86 & 1.23 \\
131227A & 582184 & 04:44:51 & 04h29m30.78s & 28d52m58.9s & 10.1 & 18.0 & -0.48 \\
131117A & 577968 & 00:34:04 & 22h09m19.36s & -31d45m44.3s & 50.9 & 11.00 & 0.27 \\
131001A & GA     & 05:37:24 & 00h33m12.96s & 25d33m25.2s & 12.4 & 4.9 & 0.96 \\
\end{tabular}
\caption{The 18 \emph{Swift}-detected GRBs that occurred in the HAWC field of view during HAWC-111 operations for which data is available. GA means the burst was found in a ground analysis.}
\label{tab1}
\end{table}

\section{Analysis Method}
The analysis described here was originally developed to run at the HAWC site (online), but is currently also used for offline data analysis. No gamma-hadron separation is applied, because the most efficient gamma-hadron separation were not available online. Since the EBL cuts of the GRB spectra at TeV energies, most of the signal is expected at lower energies, were the efficiency of the gamma-hadron separation is lowest. Therefore, at the lowest energies the gamma-hadron separation will only modestly increase the sensitivity.

The analysis consist in defining an angular bin (the search bin) around the position of the GRB, determining an estimated number of background events that will appear in the search bin and measuring the number of events above background in the bin. Besides the angular bin, the only other cut applied is that a successful angular fit of the event is required.

\begin{figure}[htbp]
\centering
\includegraphics[width=.5\textwidth]{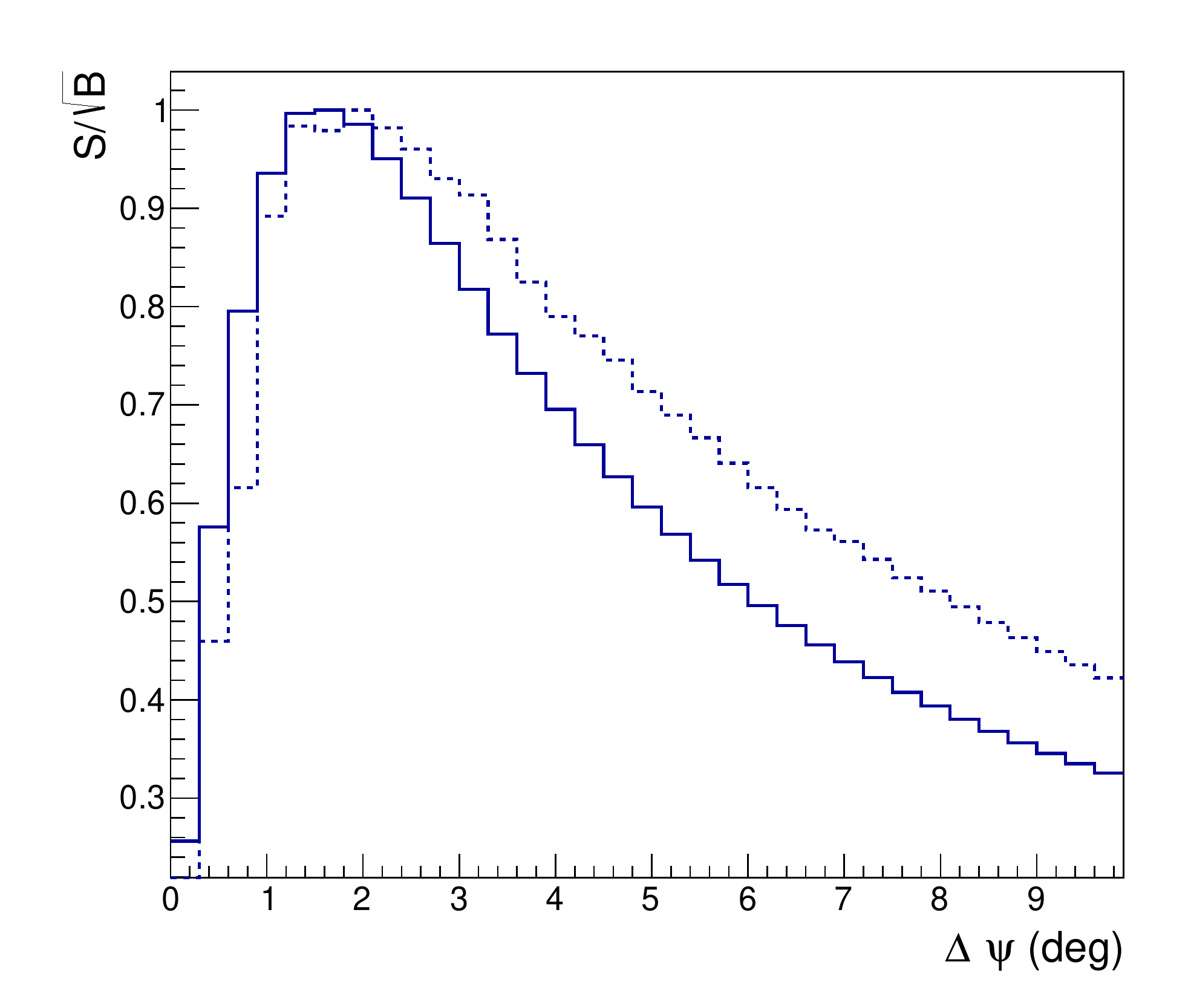}
\caption{The figure shows the signal (S) over the square root of the background (B) versus the size of the angular bin for a GRB with an $E^{-2}$ spectrum in HAWC-111. The solid (dotted) curve corresponds to a GRB at redshift 0.3 (2.0), assuming EBL attenuation according to the fiducial model in \cite{bib:EBL}. The maximum of both curves is scaled to 1 for clarity.}
\label{angular_cut_optimisation}
\end{figure}

The size of the angular bin ($\Delta\Psi$) is optimised using Monte Carlo simulations of gamma and cosmic-ray air showers and a detailed simulation of the detector. The optimisation was done for a GRB with an $E^{-2}$ spectrum, which is the typical spectrum observed by \emph{Fermi}-LAT at high energies \cite{bib:Fermi_LAT_GRB_catalogue}. The burst was simulated at various zenith angles and redshifts, modelling the absorption on the EBL according to the fiducial model in \cite{bib:EBL}. The different trigger thresholds for the HAWC-111 operation were also simulated. The optimal $\Delta\Psi$ does not depend strongly on zenith angle or trigger threshold. It does however depend on the assumed redshift, because the angular resolution of the detector depends on the number of PMTs with signal in the array, which correlates with the energy of the primary gamma ray. As the redshift increases, the signal is dominated by lower energetic gamma rays that have poorer resolution. For most GRBs, the redshift remains unknown, therefore the angular bin is conservatively chosen to be $3\arcdeg$. The optimisation of the search circle for two redshifts is shown in Fig.~\ref{angular_cut_optimisation}.

\begin{figure}[htbp]
\centering
\includegraphics[width=.7\textwidth]{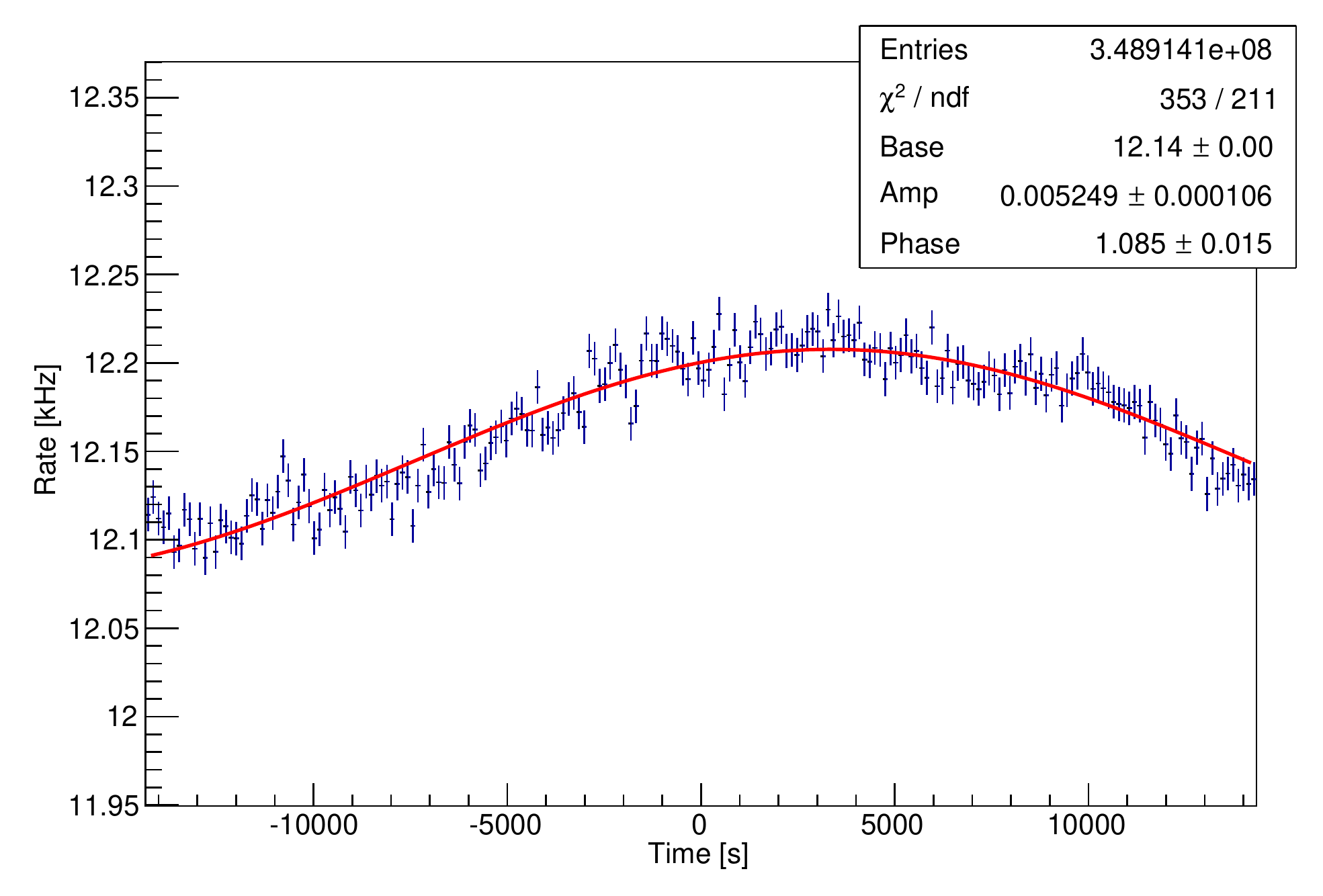}
\caption{Allsky event rate around the trigger time of GRB~140423A at zero. The data are well fitted by a constant plus a sinusoidal modulation on top. The fitted amplitude is relative to the constant.}
\label{allsky_event_rate_GRB140423A}
\end{figure}

Currently, the search duration is $T_{90}$, the central time interval, in which 90\% of the prompt \emph{Swift} Burst Alert Telescope (BAT) flux is detected, starting at the BAT trigger time. Figure~\ref{allsky_event_rate_GRB140423A} shows the allsky event rate around the time of GRB~140423A in bins of width $T_{90}$. The rate is fit with a constant plus a sinusoidal oscillation with a 12~h period on top. The sinusoidal variation of the order of $\sim0.5$\% comes from the atmospheric tide that affects the air shower propagation. The magnitude of this effect is very stable over the last two years. The distribution of excess between the data and the fit is only slightly broader than expected for a perfect Poissonian counting experiment.

\begin{figure}[htbp]
\centering
\includegraphics[width=.7\textwidth]{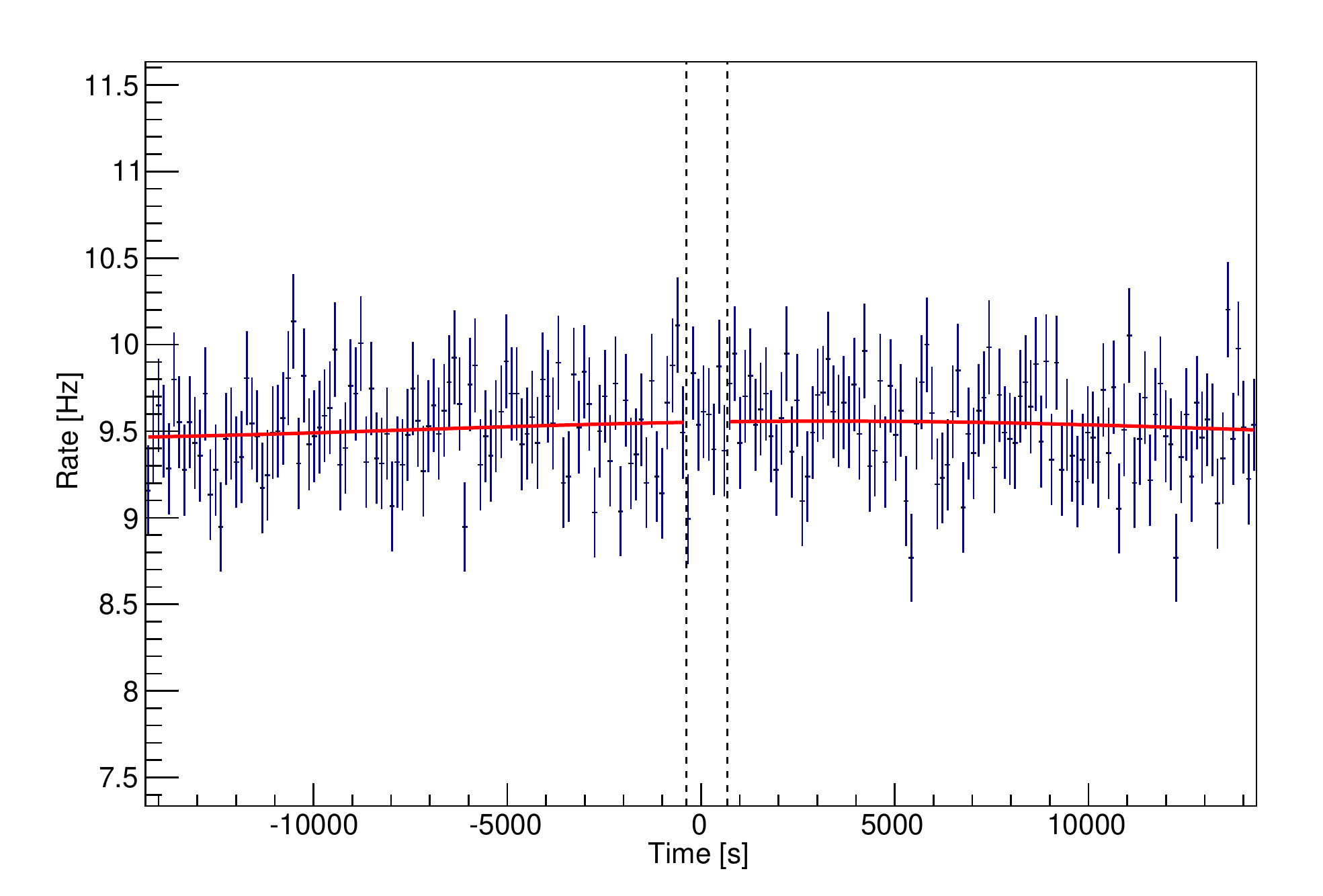}
\caption{Event rate for on-observation (at zero) and off-observations before and after the trigger time of GRB~140423A. The red line shows the background model that is created by scaling the allsky event rate fit. The horizontal dashed lines show the region where data is not used for the scaling of the background model.}
\label{on_source_event_rate_GRB140423A}
\end{figure}

The background estimate is obtained using the classical ON/OFF method (for example used in the Whipple observatory \cite{bib:ON_OFF}). In Fig.~\ref{on_source_event_rate_GRB140423A}, when the time equals zero, the search bin is centred on the GRB position and follows it for the duration of $T_{90}$ (on-observation). For all other times (off-observations), the search bin is offset in Right Ascension by multiples of $T_{90}$, thus observing an empty field of the sky that covers equal zenith angles as the on-observation. Due to the continuous data taking of HAWC, off-observations are available before and after the GRB. A background estimate at each point is obtained by taking the fit of the allsky event rate and scale it to match the summed counts in Fig.~\ref{on_source_event_rate_GRB140423A}. Counts between -380~s and 680~s are not used in the sum, because due to the size of the HAWC point spread function it takes some time before the GRB position has moved out of the search bin. It can be seen that the sinusoidal oscillation is a negligible correction in a $3\arcdeg$ search circle and fitting the off-observations with a constant yields consistent results. The counts in the on-observation are compared to the estimate from the background model obtained using the off-observations and the significance is calculated assuming a Poisson distribution.

\section{Results}
\begin{figure}[htbp]
\centering
\includegraphics[width=.7\textwidth]{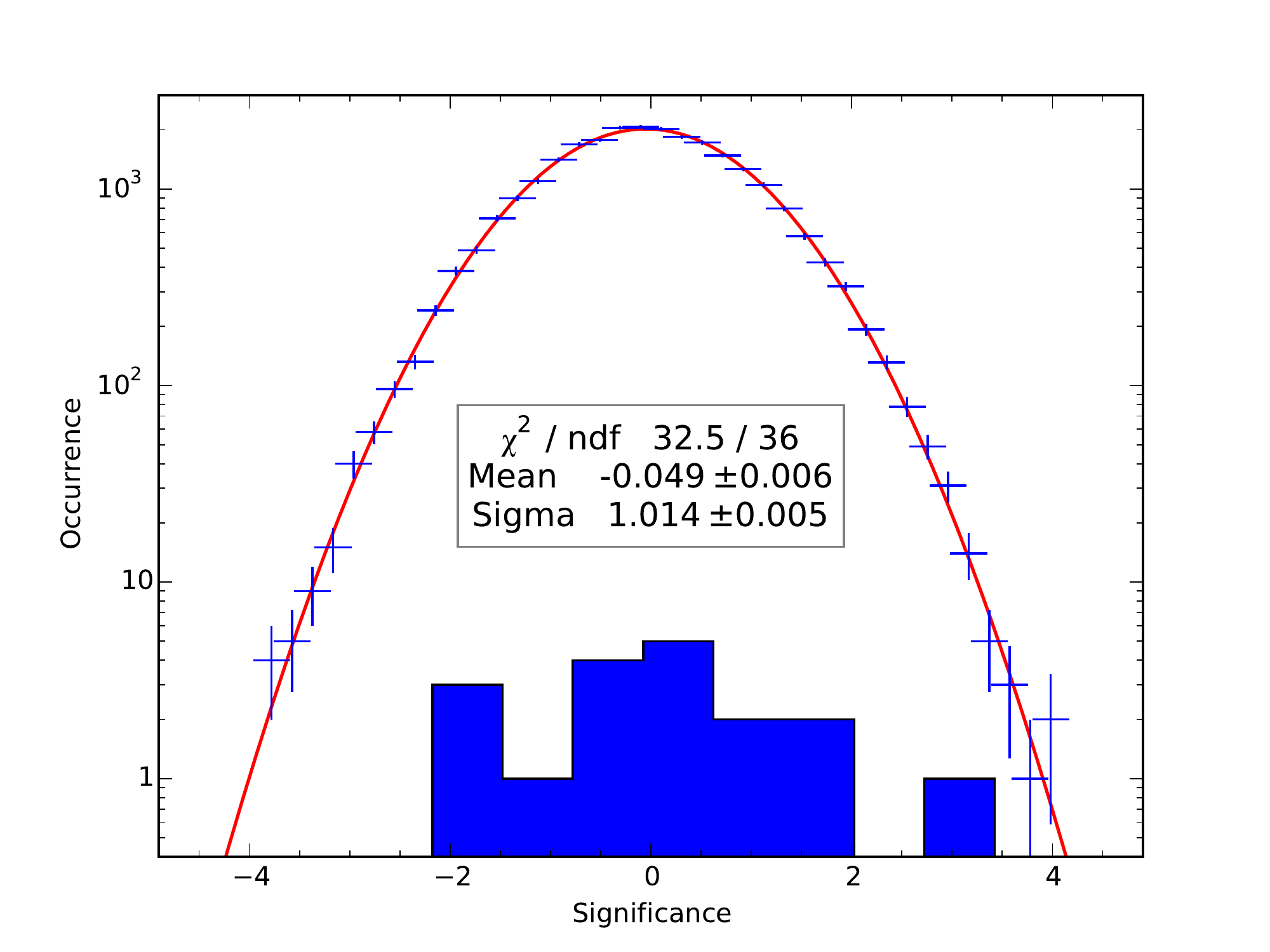}
\caption{The blue histogram shows the significance distribution obtained from analysing the 18 \emph{Swift}-detected GRBs in Table~\protect\ref{tab1}. The blue points show the significance distribution that is obtained by applying the analysis method to 25,188 fake GRB positions. The results of the Gaussian fit (red line) are given in the inset box.}
\label{significance_distribution}
\end{figure}

Fig.~\ref{significance_distribution} shows the significance distribution that is obtained by applying the analysis method described above to 25,188 fake GRB positions in HAWC-111 data, which are placed at zenith angles between 0\arcdeg and 60\arcdeg. The distribution is very well fitted by a Gaussian, showing that the background is well understood to the 4~$\sigma$ level.

Applying the analysis method to the 18 \emph{Swift}-detected GRBs from Table~\ref{tab1} gives the blue histogram of significances in Fig.~\ref{significance_distribution} (the individual significance for each burst are listed in Table~\ref{tab1}). GRB~140607A is the only burst with a significance above 3~$\sigma$. Accounting for the trial factor of looking at 18 bursts at once, the significance drops to 2.5~$\sigma$. GRB~140607A was detected by the BAT only, because due to a Sun observing constraint, \emph{Swift} couldn't slew to the BAT position. No other observations were reported for this GRB.

\section{Outlook}
In this contribution initial results for GRBs in the HAWC field of view during HAWC-111 were shown. None of the GRBs is significant above 3~$\sigma$ when accounting for trial factors. An alternative self-triggered search for GRBs is also planned for HAWC \cite{bib:Josh-ICRC-2015}. This search has a worse detection threshold than the one presented here, but it may find bursts not reported by satellites.

\emph{Fermi}-LAT found that the $>100\nobreakspace\rm{MeV}$ emission of GRBs is temporally extended and starts later than the keV--MeV emission \cite{bib:Fermi_LAT_GRB_catalogue}. This implies that time windows other than the BAT $T_{90}$ should be searched for VHE emission as well. The angular bin size has been optimised for the next stage of the detector, called HAWC-250, and the application of gamma-hadron separation cuts for both online and offline analysis are being explored. Gamma-hadron separation is most efficient at $>$TeV energies, but the application of cuts on HAWC-250 can still improve the detection significance by 50\% to 100\%. Results of offline and online analysis of HAWC-250 and the full HAWC array will be presented elsewhere.

It has been show elsewhere that HAWC is sensitive enough to detect several historical GRBs (like GRB~090510 and 090902B) if their emission extends only slightly beyond the highest energy observed by \emph{Fermi}-LAT \cite{bib:HAWC_GRBs}. Furthermore, a GRB similar to 130427A would be detected even if the spectrum doesn't extend beyond what was observed by the LAT \cite{bib:HAWC_GRB130427A}. In addition to exceptional bursts, HAWC might detect other GRBs with a rate as high as 1--2 GRBs per year \cite{bib:HAWC_GRB_rate}. Even the absence of detections will be useful in constraining the highest energy of GRB spectra.

\section*{Acknowledgments}
\footnotesize{
We acknowledge the support from: the US National Science Foundation (NSF);
the US Department of Energy Office of High-Energy Physics;
the Laboratory Directed Research and Development (LDRD) program of
Los Alamos National Laboratory; Consejo Nacional de Ciencia y Tecnolog\'{\i}a (CONACyT),
Mexico (grants 260378, 55155, 105666, 122331, 132197, 167281, 167733);
Red de F\'{\i}sica de Altas Energ\'{\i}as, Mexico;
DGAPA-UNAM (grants IG100414-3, IN108713,  IN121309, IN115409, IN111315);
VIEP-BUAP (grant 161-EXC-2011);
the University of Wisconsin Alumni Research Foundation;
the Institute of Geophysics, Planetary Physics, and Signatures at Los Alamos National Laboratory;
the Luc Binette Foundation UNAM Postdoctoral Fellowship program.
}

\bibliography{icrc2015-0237}

\end{document}